\documentclass[twocolumn,twoside,slac_two,floatfix]{revtex4}
\usepackage{graphicx}
\usepackage{fancyhdr}
\newcommand\het{{\small HET}}
\newcommand\lrs{{\small LRS}}
\newcommand\apo{{\small APO}}
\newcommand\arc{{\small ARC}}
\newcommand\dis{{\small DIS}}
\newcommand\nmsu{{\small NMSU}}
\newcommand\sdss{{\small SDSS}}
\newcommand\cfht{{\small CFHT}}
\newcommand\eso{{\small ESO}}
\newcommand\mdm{{\small MDM}}
\newcommand\essence{{\small ESSENCE}}
\newcommand\sn{{\small SN}}
\newcommand\sne{{\small SN}e}
\newcommand\agn{{\small AGN}}
\newcommand\qso{{\small QSO}}
\newcommand\ra{{\small RA}}
\newcommand\ugriz{{\it ugriz}}
\newcommand\psf{{\small PSF}}
\newcommand\dr{{\small DR}}

\newcommand\dm{\Delta m_{15}(B)}
\newcommand\iraf{{\small IRAF}}
\newcommand\rvsao{{\small RVSAO}}
\newcommand\lax{\>\vcenter{\hbox{$<$\hskip-.75em\lower1.0ex\hbox{$\sim$}}}\>}
\newcommand\gax{\>\vcenter{\hbox{$>$\hskip-.75em\lower1.0ex\hbox{$\sim$}}}\>}

\newcommand\fsn{{\footnotesize SN}}
\newcommand\fsne{{\footnotesize SN}e}
\newcommand\fhet{{\footnotesize HET}}

\newcommand\farc{{\footnotesize ARC}}
\newcommand\fdis{{\footnotesize DIS}}
\newcommand\fnmsu{{\footnotesize NMSU}}
\newcommand\fsdss{{\footnotesize SDSS}}

\begin{document}

\title{The Fall 2004 SDSS Supernova Survey}

\author{Masao Sako (KIPAC/Stanford), Roger Romani (Stanford), Josh Frieman
  (Fermilab/U. Chicago), Jen Adelman-McCarthy (Fermilab), Andrew Becker
  (U. Washington), Fritz DeJongh (Fermilab), Ben Dilday (U. Chicago), Juan
  Estrada (Fermilab), John Hendry (Fermilab), Jon Holtzman (NMSU), Jared
  Kaplan (Stanford), Rick Kessler (U. Chicago), Hubert Lampeitl (Fermilab),
  John Marriner (Fermilab), Gajus Miknaitis (U. Washington), Adam Riess
  (STScI), Douglas Tucker (Fermilab), John Barentine (APO), Roger Blandford
  (KIPAC/Stanford), Howard Brewington (APO), Jack Dembicky (APO), Mike
  Harvanek (APO), Suzanne Hawley (U. Washington), Craig Hogan (U. Washington),
  David Johnston (Princeton), Steve Kahn (KIPAC/Stanford), Bill Ketzeback
  (APO), Scot Kleinman (APO), Jerzy Krzesinski (APO), Dennis Lamenti (SFSU),
  Dan Long (APO), Russet McMillan (APO), Peter Newman (APO), Atsuko Nitta
  (APO), Robert Nichol (Portsmouth), Ryan Scranton (U. Pittsburgh), Erin
  Sheldon (U. Chicago), Stephanie Snedden (APO), Chris Stoughton (Fermilab),
  Don York (U. Chicago), and the SDSS Collaboration}

\begin{abstract}

In preparation for the Supernova Survey of the Sloan Digital Sky Survey
(\fsdss) II, a proposed 3-year extension to the \fsdss, we have conducted an
early engineering and science run during the fall of 2004, which consisted of
approximately 20 scheduled nights of repeated imaging of half of the southern
equatorial stripe.  Transient supernova-like events were detected in near
real-time and photometric measurements were made in the five \fsdss\ filter
bandpasses with a cadence of $\sim 2$~days.  Candidate type Ia supernovae
(\fsne) were pre-selected based on their colors, light curve shape, and the
properties of the host galaxy.  Follow-up spectroscopic observations were
performed with the Astrophysical Research Consortium 3.5m telescope and the
9.2m Hobby-Eberly Telescope to confirm their types and measure the redshifts.
The 2004 campaign resulted in 22 spectroscopically confirmed \fsne, which
includes 16 type Ia, 5 type II, and 1 type Ib/c.  These \fsn\ Ia will help
fill in the sparsely sampled redshift interval of $z = 0.05 - 0.35$, the
so-called 'redshift desert', in the Hubble diagram.  Detailed investigation of
the spectral properties of these moderate-redshift \fsne\ Ia will also provide
a bridge between local \fsne\ and high-redshift objects, and will help us
understand the systematics for future cosmological applications that require
high photometric precision.  Finally, the large survey volume also provides
the opportunity to select unusual supernovae for spectroscopic study that are
poorly sampled in other surveys.  We report on some of the early results from
this program and discuss potential future applications.

\end{abstract}

\maketitle

\thispagestyle{fancy}

\section{Introduction}

One of the three primary scientific components of \sdss\ II, a proposed
three-year extension to the \sdss, is a time domain study, which involves
repeated imaging of the southern equatorial stripe for three three-month
periods during the Fall months of 2005 -- 2007.  The Supernova Survey is a
major subcomponent of this study, and its primary goal is to obtain
high-quality, multi-color light curves of $\sim 200$ \sn\ Ia in the
intermediate redshift interval of $z = 0.05 - 0.35$.  This complements
existing searches for \sne\ at low redshifts (e.g., Lick Observatory Supernova
Search, Carnegie Supernova Program, Nearby Supernova Factory) and those at
high redshifts (e.g., \cfht\ Supernova Legacy Survey, \essence), and will not
be targeted by other telescopes in the coming years.  The \sdss\ 2.5m
telescope is particularly well-suited for this type of search with its ability
to survey large areas of the sky to moderate depths in several filters all in
a single night of observing.  In addition, it allows us to take advantage of
one of the most well-calibrated photometric systems that is currently
available, which is crucial to help understand and quantify the systematic
uncertainties associated with \sne\ Ia as cosmological distance indicators.

\begin{table}[ht]
\begin{center}
\caption{List of \fsdss\ 2.5m imaging scans of strip 82 N for the Fall 2004
  campaign}
\vspace{0.3cm}
\begin{tabular*}{50mm}{@{\extracolsep{\fill}}cccr@{\hspace{0.4cm}--\hspace{0.4cm}}l}
\hline \textbf{Date} & \textbf{run} & & \multicolumn{2}{c}{\textbf{RA Range (hrs)}}
 \\ \hline
 9/22  & 4894 & &   $19.62$ & $4.90$     \\
 9/24  & 4895 & &   $20.69$ & $3.98$     \\
 10/8  & 4896 & &   $22.08$ & $4.14$     \\
 10/10 & 4874 & &   $19.95$ & $5.82$     \\
 10/16 & 4894/5 & & $21.88$ & $4.64$     \\
 10/18 & 4899 & &   $0.66$ & $4.14$     \\
 10/20 & 4805 & &   $0.14$ & $4.73$     \\
 10/24 & 4917 & &   $19.74$ & $3.47$     \\
 11/3  & 4927 & &   $20.64$ & $4.12$     \\
 11/5  & 4933 & &   $20.53$ & $4.16$     \\
 11/10 & 4948 & &   $1.27$ & $4.11$     \\ \hline
\end{tabular*}
\label{run_list}
\end{center}
\end{table}

In preparation for \sdss\ II, we have conducted a pilot study during the Fall
of 2004 using the \sdss\ 2.5m imaging telescope.  This run consisted of
repeated scans of half of the southern equatorial stripe over a period of
$\sim 1.5$~months scheduled with a cadence of approximately 2 days.  The area
covered is approximately half of that planned for \sdss\ II and the duration
is $\sim 6$~times shorter than the total duration over three years, which
makes this pilot run $\sim 1/12$ of the size of \sdss\ II.  We report here
some of the early results from this program, which resulted in 22
spectroscopically identified \sne\ and light curves of an additional $\sim
2$~dozen unconfirmed events.  We also briefly discuss our plans for \sdss\ II.
Although the primary goal of the Supernova Survey is to study the properties
of intermediate-redshift type Ia \sne, its large survey volume will enable us
to find a large number of other types of \sne\ including peculiar ones as
well.  We briefly comment on a few potential applications.

\section{Observations}
\subsection{SDSS Imaging and Target Selection}

\begin{figure}[tb]
\centering
\includegraphics[width=40mm]{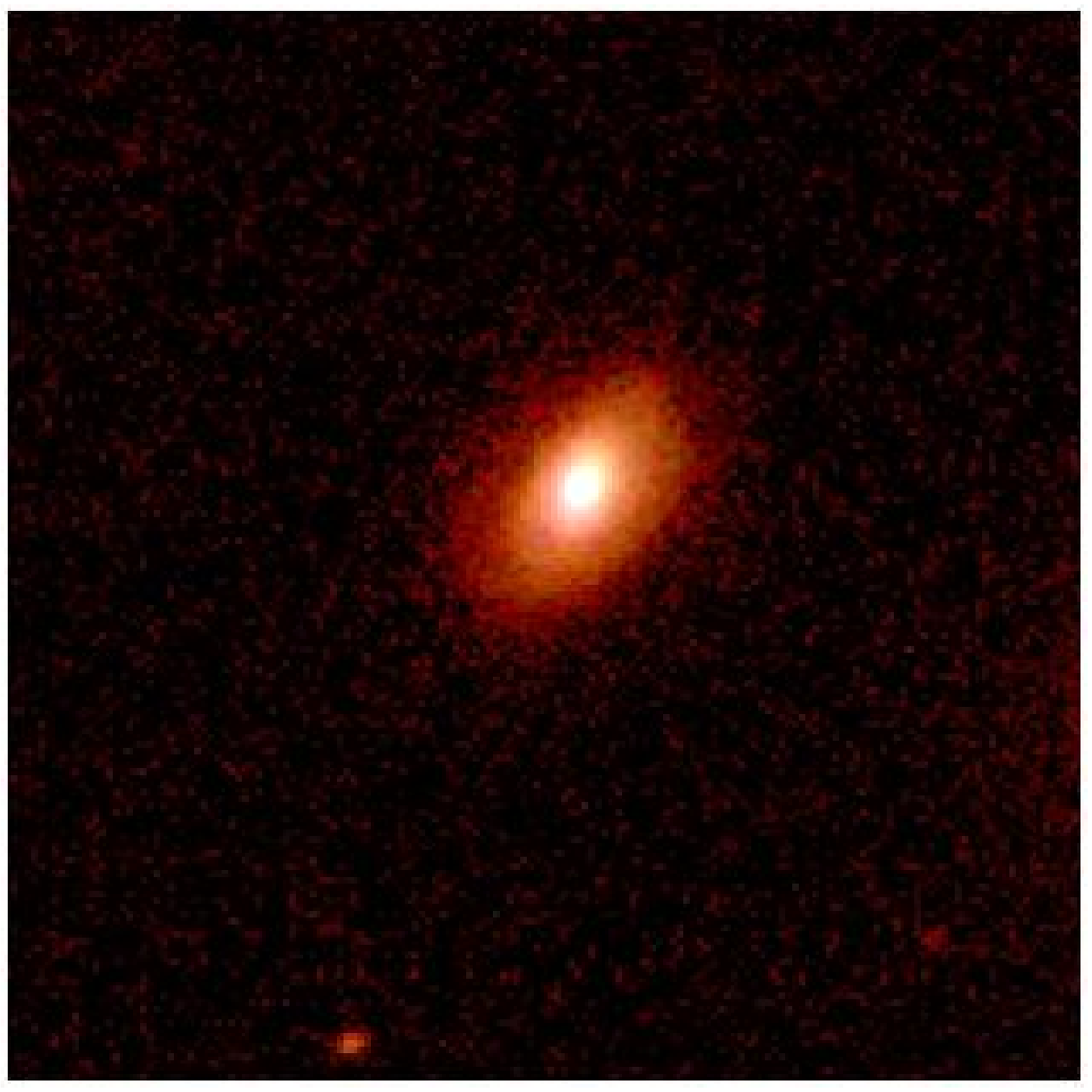}
\includegraphics[width=40mm]{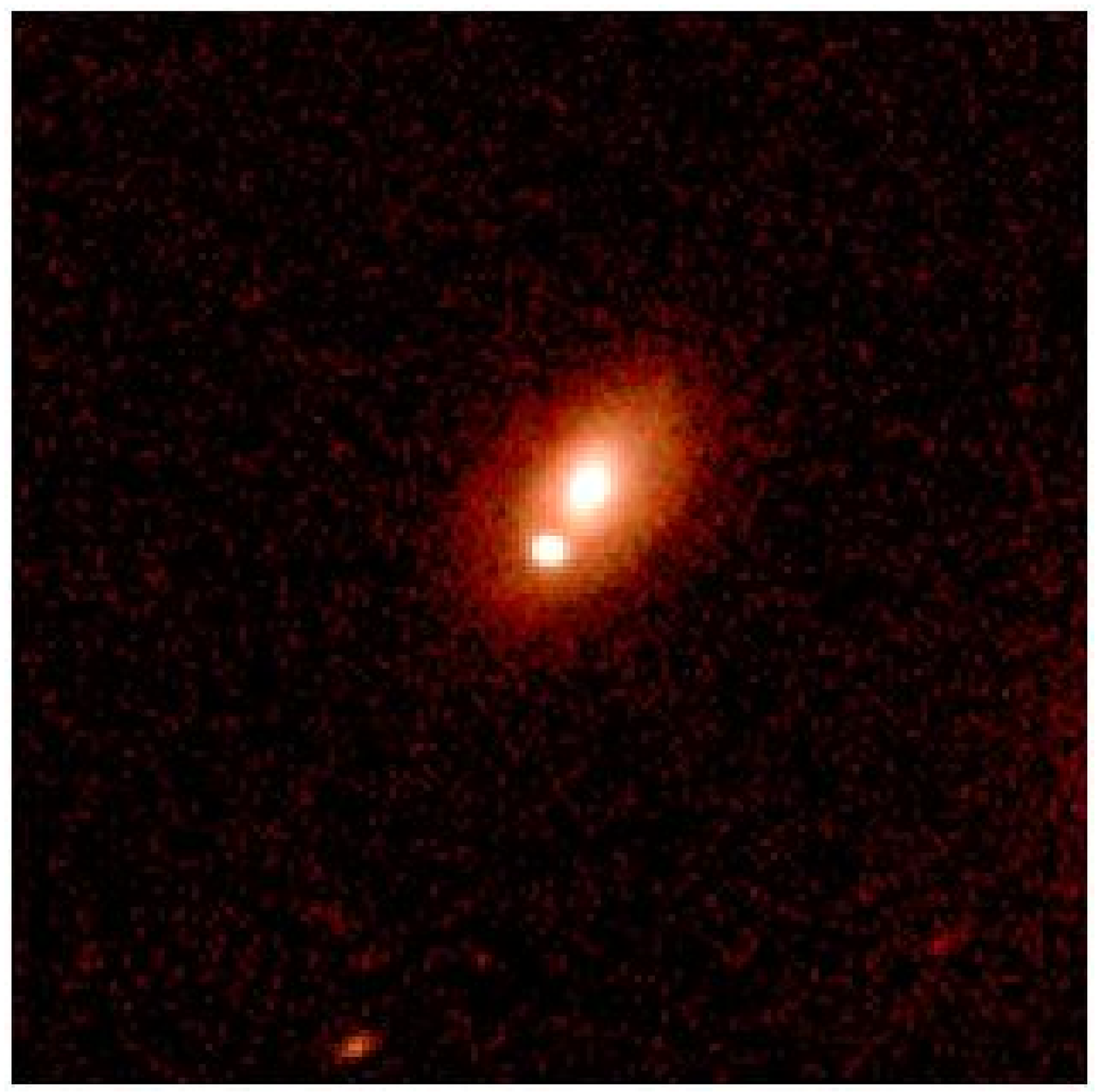} \\
\includegraphics[width=40mm]{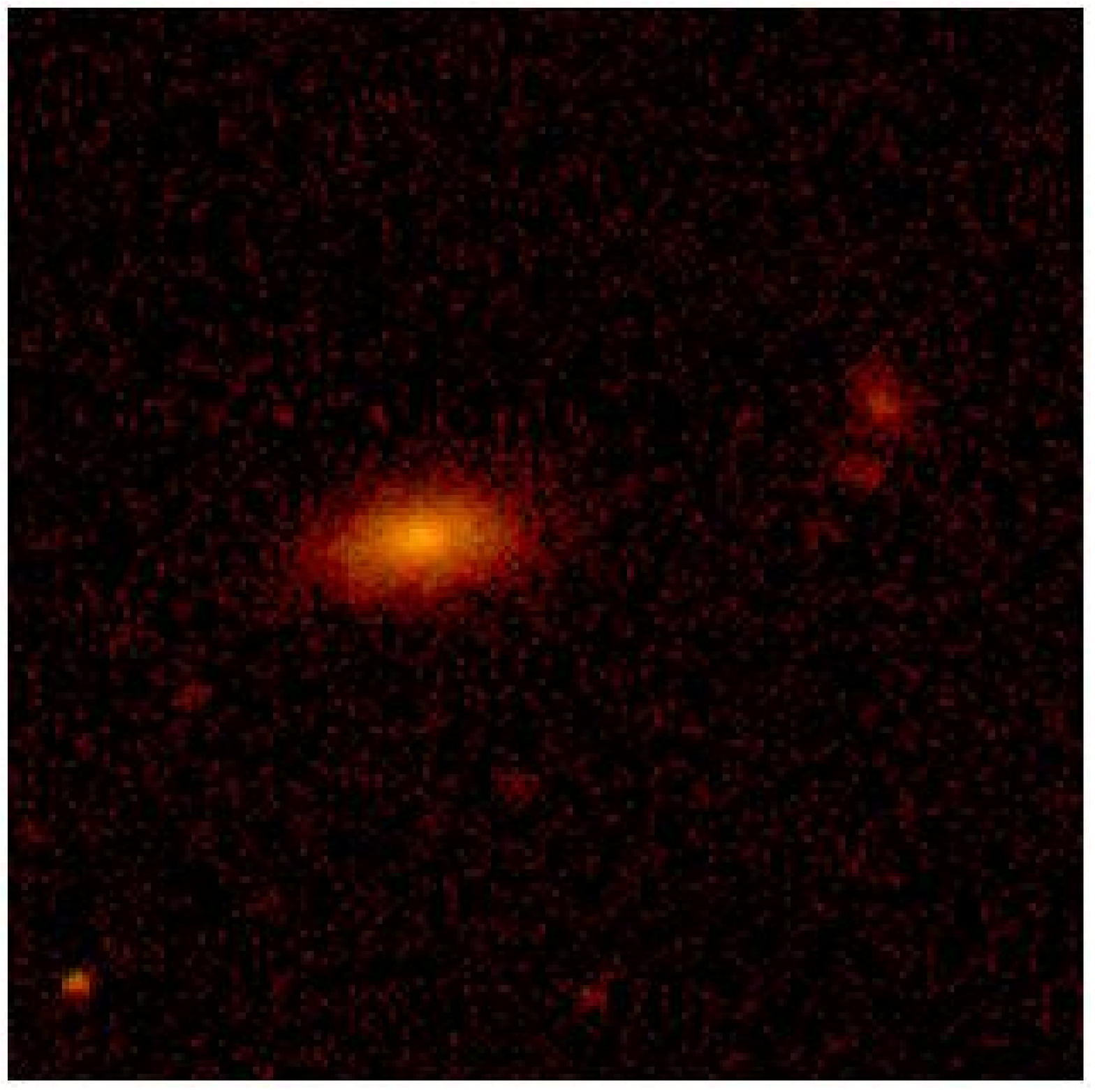}
\includegraphics[width=40mm]{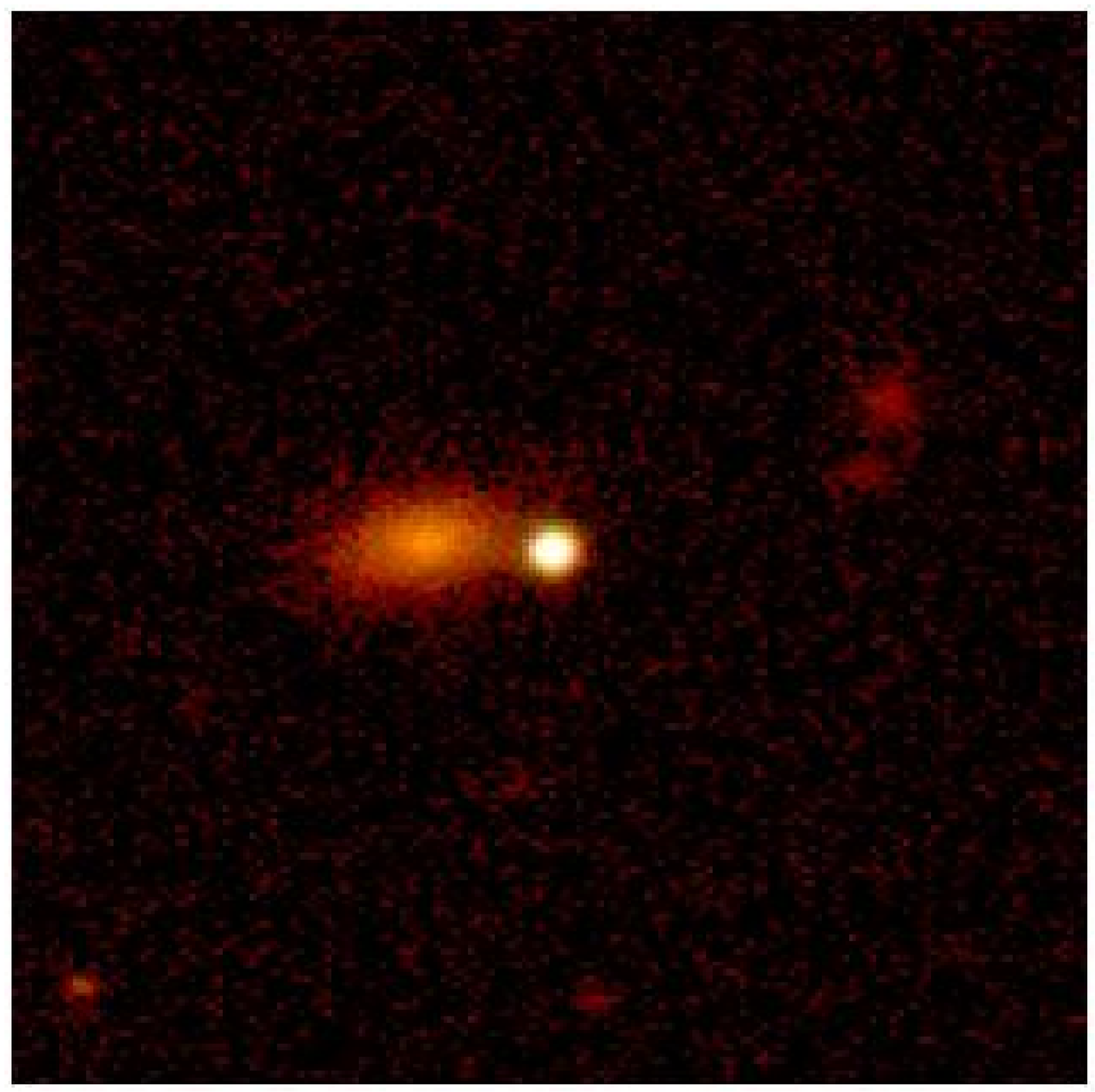} \\
\includegraphics[width=40mm]{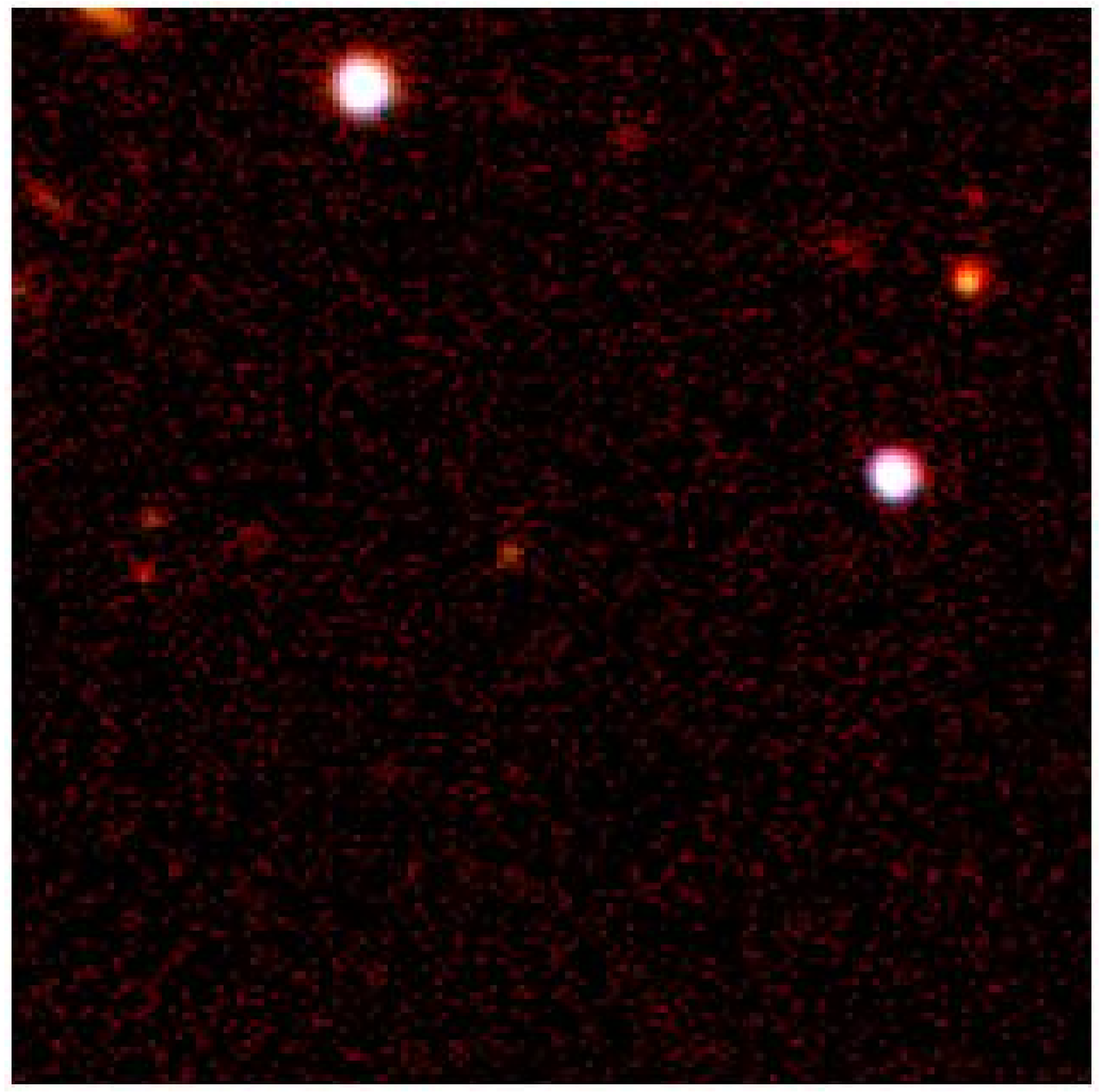}
\includegraphics[width=40mm]{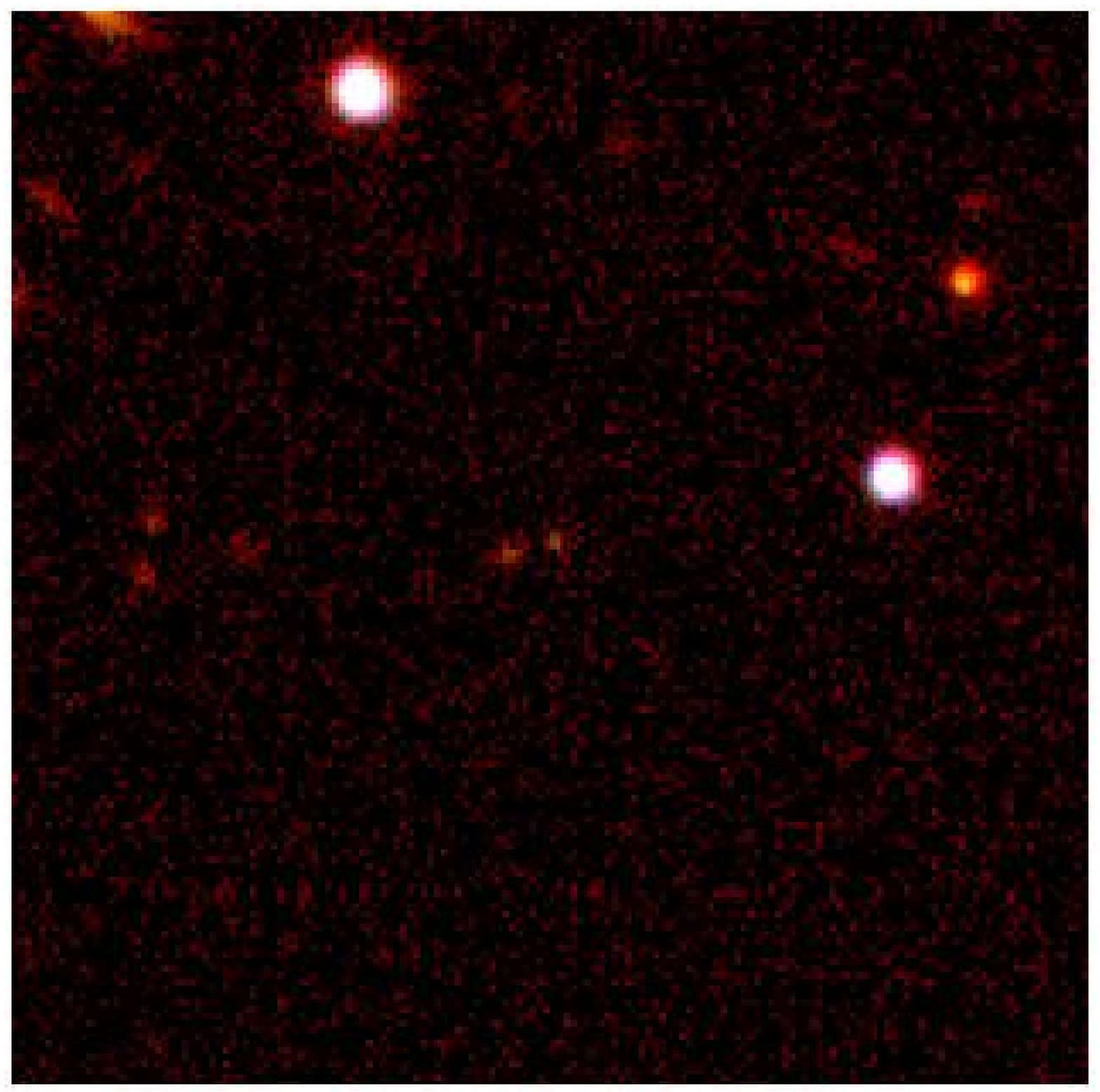} \\
\caption{\fsdss\ images of three type Ia \sne\ discovered and
  spectroscopically confirmed during the 2004 campaign -- from top to bottom,
  \fsn 2004hu ($z = 0.0487$), \fsn 2004ie ($z = 0.0513$), and \fsn 2004if ($z
  = 0.322$).  The left panels show the template images before the supernova,
  which are shown on the right panels.  The images are 1$^\prime \times
  1^\prime$ and are all centered on the locations of the
  supernovae.} \label{snia_images}
\end{figure}

\begin{figure}[tb]
\centering
\includegraphics[width=40mm]{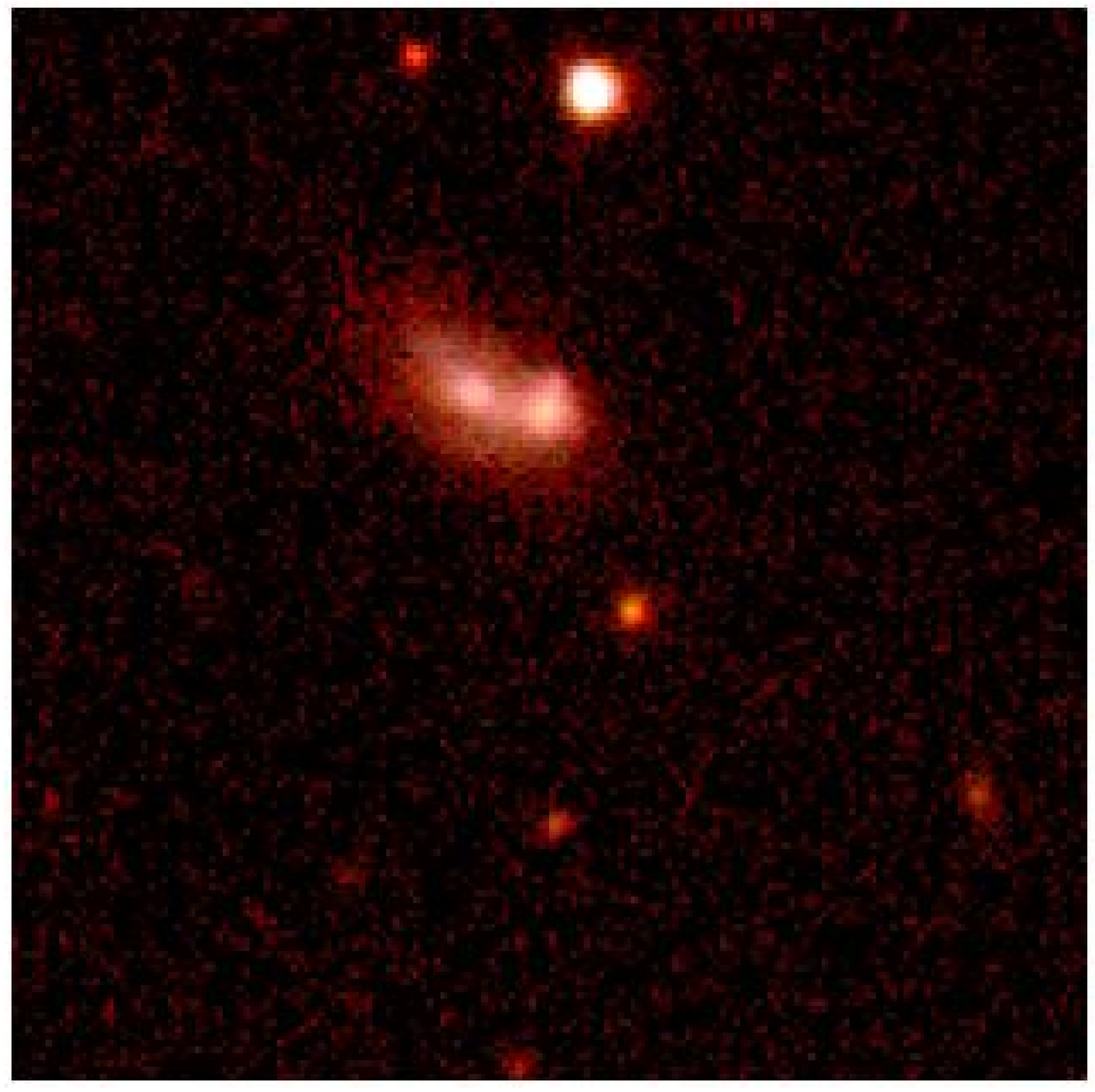}
\includegraphics[width=40mm]{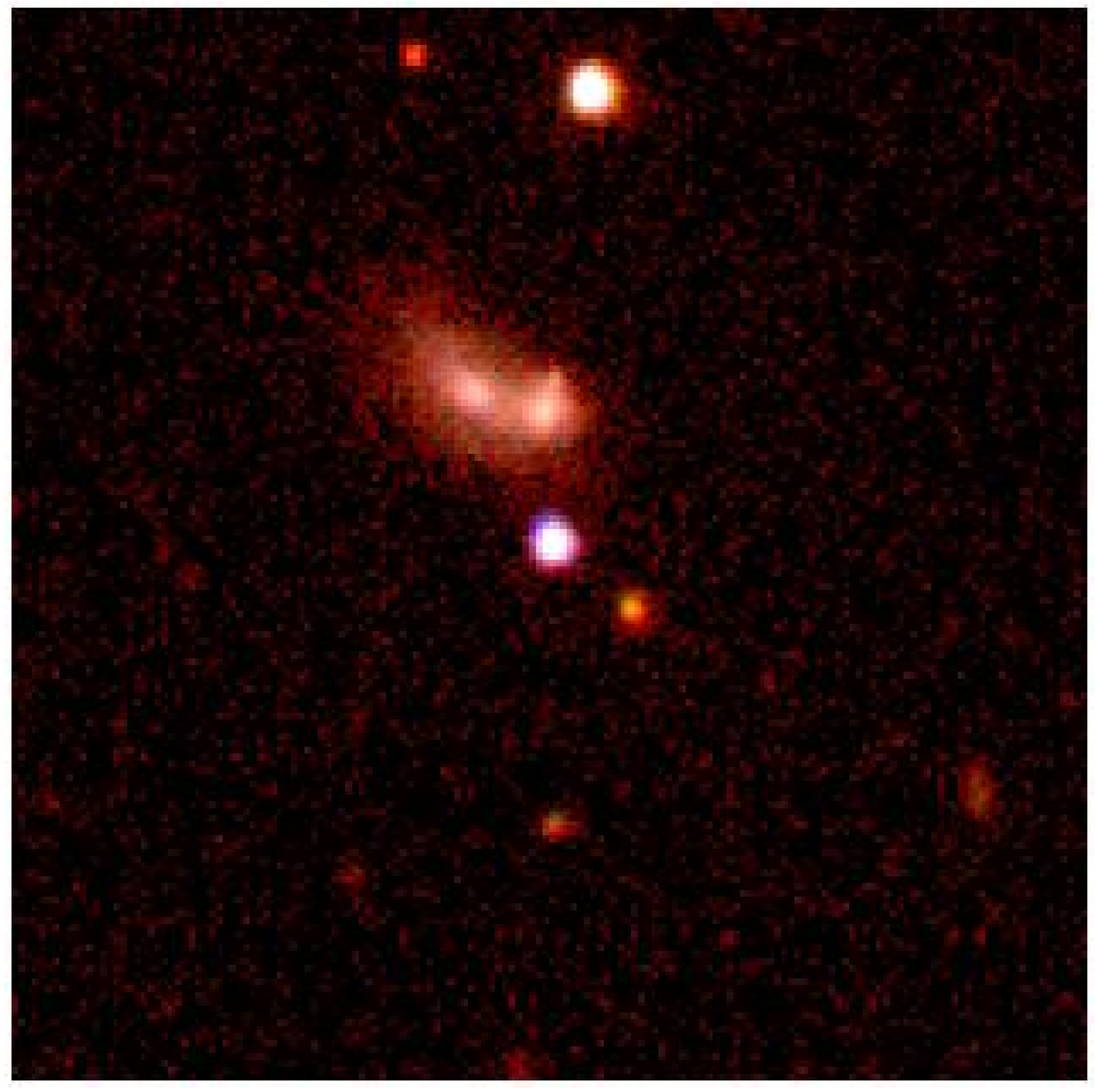} \\
\includegraphics[width=40mm]{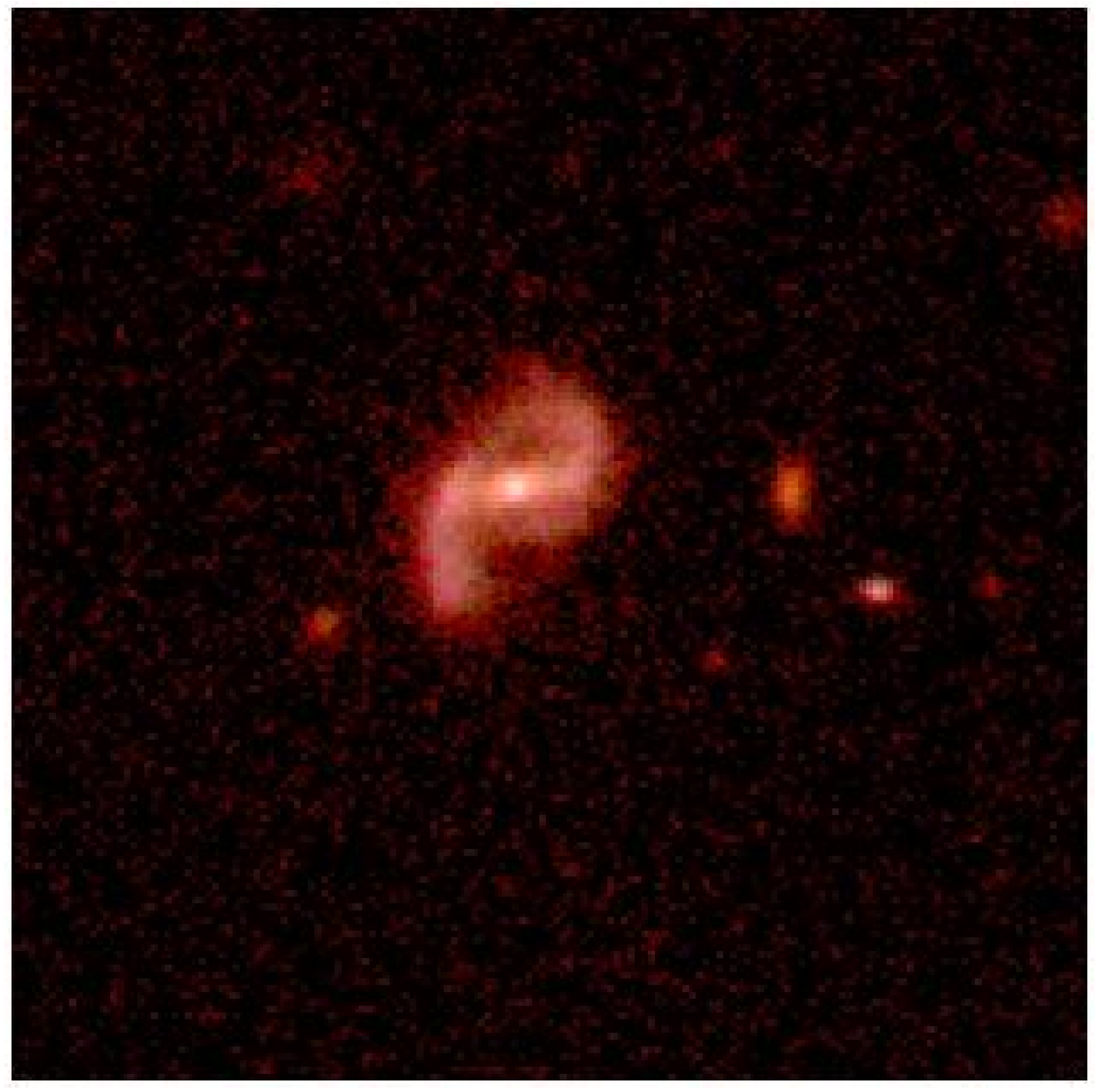}
\includegraphics[width=40mm]{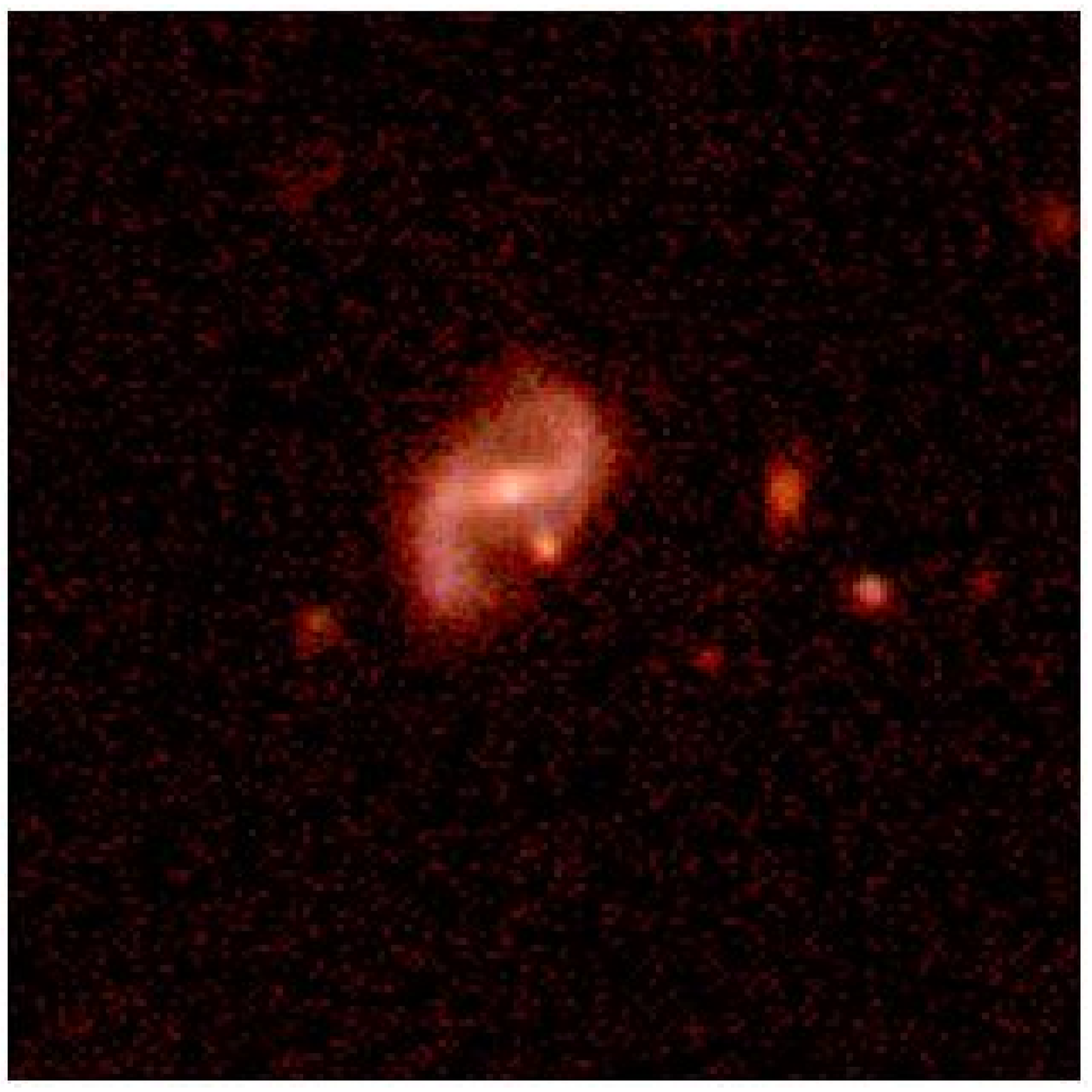} \\
\includegraphics[width=40mm]{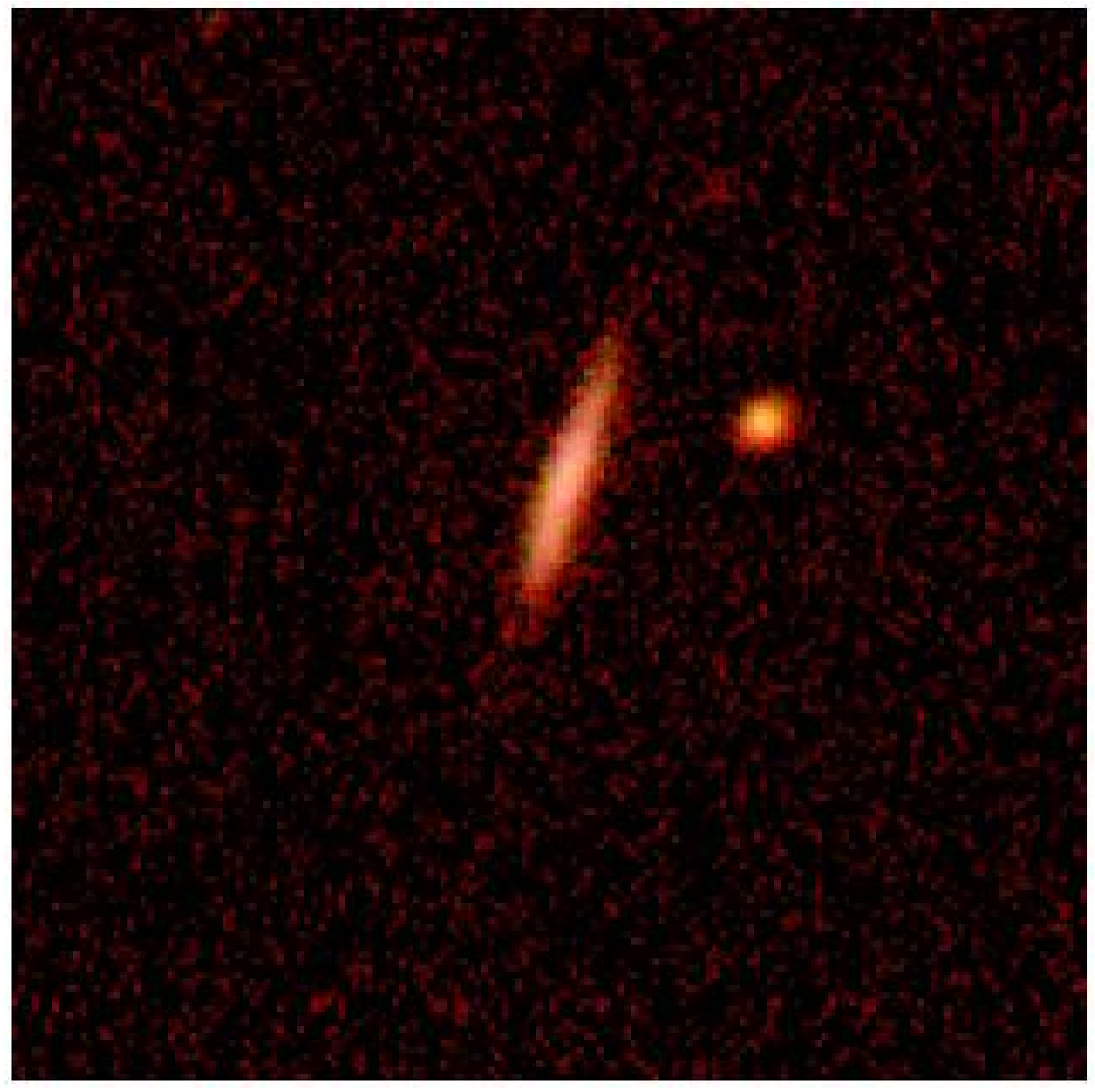}
\includegraphics[width=40mm]{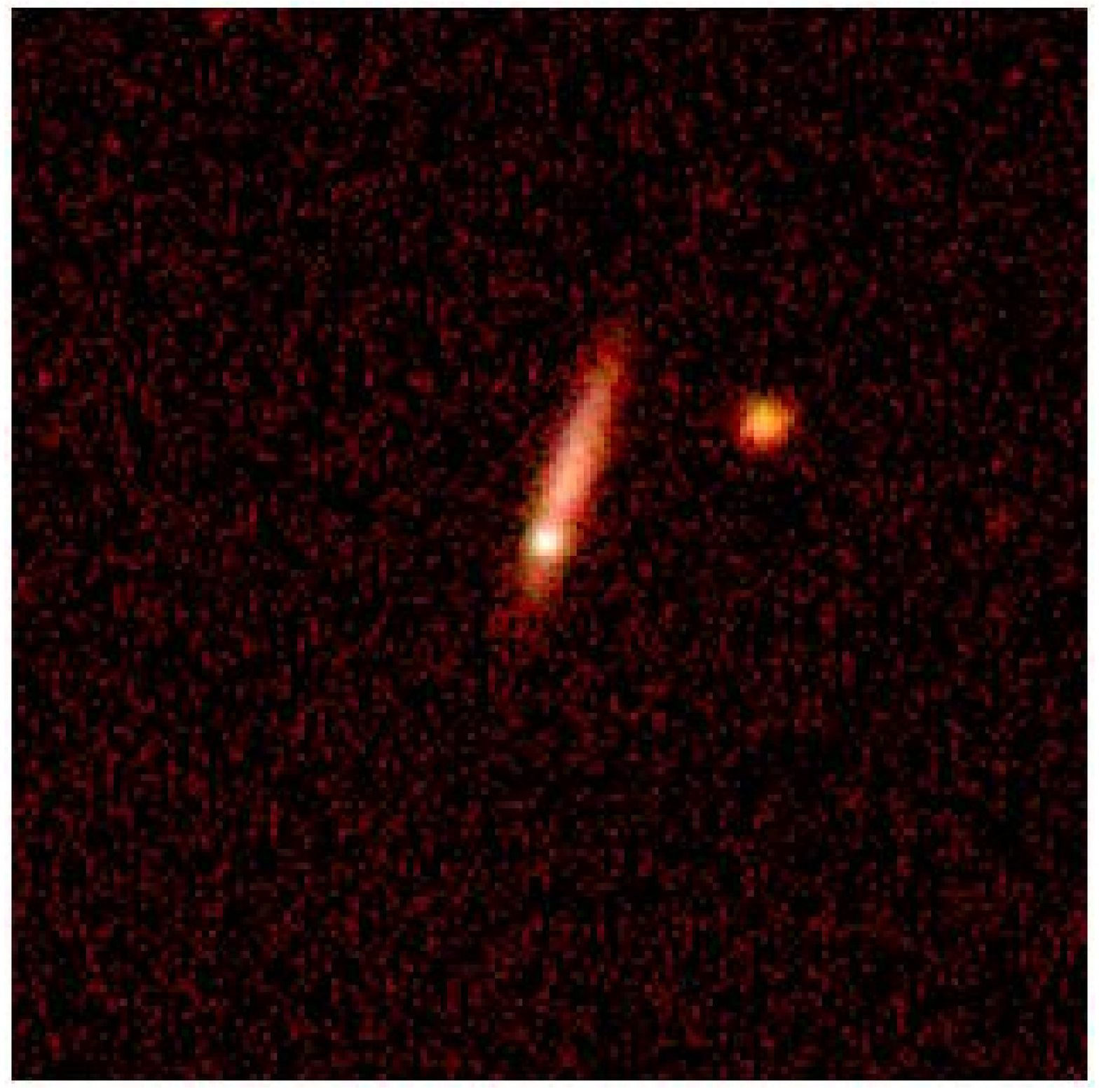} \\
\caption{Same as in Figure~\ref{snia_images} for, from top to bottom, \fsn
  2004hx (type II; $z = 0.0143$)), \fsn 2004ht (type II; $z = 0.0668$), and
  \fsn 2004ib (type Ib/c; $z = 0.0565$).} \label{snother_images}
\end{figure}

\sdss\ imaging observations of half of the southern equatorial stripe (strip
82 N) were planned for approximately 20 nights between Sept. 21 and Nov. 15,
2004.  This region covers approximately 150 square degrees of the sky, and we
were able to obtain nearly complete scans of the strip on 6 nights and partial
scans during another 5 nights (Table~\ref{run_list}).  Note the presence of
relatively long gaps during 9/24 -- 10/8 and 10/24 -- 11/3, which were mostly
bright time.  The ranges in \ra\ shown are only approximate.

Processed \ugriz\ images were generated in near real-time at Apache Point
Observatory (\apo) and the $g$ and $r$ frames were run through a pipeline
software that searches for transient events through image subtraction.  During
this run, we considered sources that were detected with a statistical
significance of at least $\sim 10~\sigma$ in the subtracted images of both the
$g$ and $r$ bands for manual inspection.  This relatively high threshold
naturally biased our selection towards lower redshift \sne\ and to those that
are well-separated from the cores of bright host galaxies.  An additional
requirement of positional matches in the $g$ and $r$ images, whose exposures
are separated by $\sim 5$~minutes, removed most fast-moving asteroids.
Matches with previously catalogued variable objects (stars and \agn) and with
bright stars were vetoed in the software as well.  A few \sne\ that are easily
visible in the raw images are shown in Figures~\ref{snia_images} and
\ref{snother_images}.

Objects that made it through the automated selection were then inspected one
by one and prioritized according to their likeliness to be \sne.  This was
done after every imaging run with the \sdss\ 2.5m.  Since only the $g$ and $r$
bands were processed automatically though image subtraction, we did not make
full use of the color information for initial photometric typing of the
events.  In general, objects that appear in the differenced image near the
core of the host galaxy are ranked low, since they are relatively more likely
to be \agn s.  Unresolved point-like sources that are isolated from any
potential host galaxy are likely to be variable stars or \qso s, and were also
given low priority.  All of the remaining events are potential \sne.  For
\sdss\ II, all images will be processed automatically through image
subtraction in $g$, $r$, and $i$.  In addition, resulting frames containing
potentially interesting objects will be rapidly processed through image
subtraction in $u$ and $z$ as well, allowing us to make use of the full color
information for photometric typing of candidate \sne.

\begin{table}[ht]
\begin{center}
\caption{List of Spectroscopically Confirmed SNe from the Fall 2004 Campaign
  \cite{sne_iauc}}
\vspace{0.3cm}
\begin{tabular}{ccccc}
\hline \textbf{IAUC ID} & \textbf{SDSS ID} & \textbf{Type} & \textbf{Redshift}
& \textbf{Telescope} \\ \hline
 2004ht & SN24  & II   & 0.0668 & ARC/HET \\
 2004hu & SN15  & Ia   & 0.0487 & ARC     \\
 2004hv & SN9   & II   & 0.0613 & ARC     \\
 2004hw & SN10  & Ia   & 0.0601 & HET     \\
 2004hx & SN18  & II   & 0.0143 & ARC     \\
 2004hy & SN12  & II   & 0.0587 & ARC     \\
 2004hz & SN5   & Ia   & 0.1427 & HET     \\
 2004ia & SN30  & Ia   & 0.1431 & ARC     \\
 2004ib & SN20  & Ib/c & 0.0565 & HET     \\
 2004ic & SN111 & II   & 0.093  & ARC     \\
 2004id & SN19  & Ia   & 0.1444 & ARC     \\
 2004ie & SN83  & Ia   & 0.0513 & ARC     \\
 2004if & SN25  & Ia   & 0.322  & HET     \\
 2004ig & SN133 & Ia   & 0.1831 & ARC     \\
 2004ih & SN128 & Ia   & 0.1538 & ARC     \\
 2004ii & SN172 & Ia   & 0.1973 & ARC     \\
 2004ij & SN247 & Ia   & 0.218  & HET     \\
 2004ik & SN242 & Ia   & 0.1691 & HET     \\
 2004il & SN191 & Ia   & 0.108  & ARC     \\
 2004im & SN176 & Ia   & 0.134  & ARC     \\
 2004in & SN194 & Ia   & 0.1612 & ARC     \\
 2004io & SN171 & Ia   & 0.1502 & ARC     \\ \hline
\end{tabular}
\label{sn_list}
\end{center}
\end{table}

\subsection{Additional Imaging Observations}

For a handful of the spectroscopically confirmed \sne, additional imaging
observations were performed with the New Mexico State University (\nmsu) 1m
and the Astrophysical Research Consortium (\arc) 3.5m telescopes at \apo.  The
images were taken with the \sdss\ $g$, $r$, and $i$ filters whenever possible.
The main objectives of these observations were to obtain more densely-sampled
light curves on a handful of candidates and to provide additional photometric
points beyond the survey duration especially for those events that were
discovered towards the end of the run.  This was especially important for the
test run, since its 1.5-month duration was not optimal for harvesting full
\sne\ light curves.  Since all of these telescopes are at \apo, however, they
could not be used as replacements to fill in light curve points during poor
weather conditions.

\begin{figure}[htb]
\centering
\includegraphics[width=80mm]{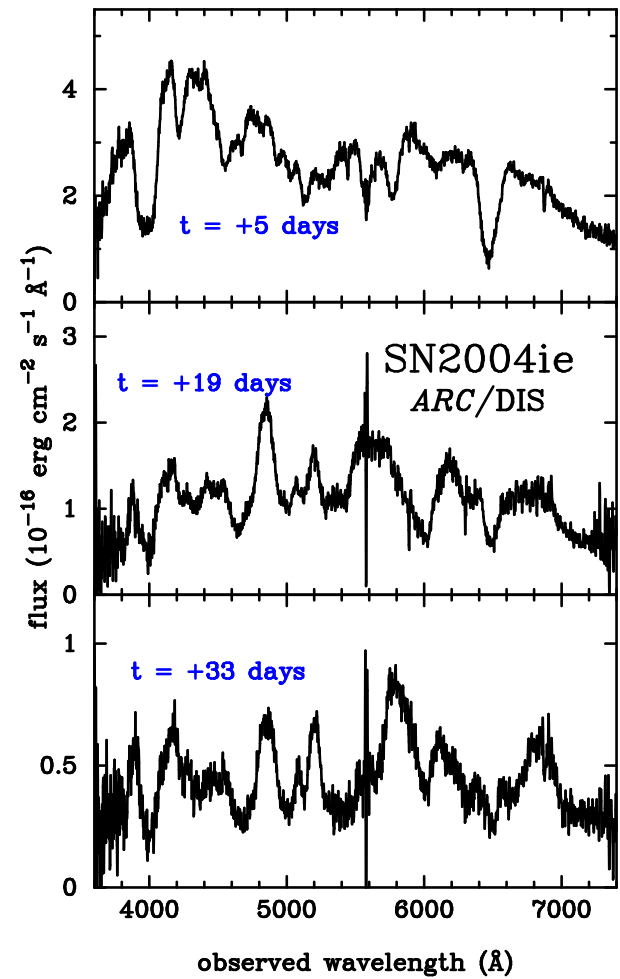}
\caption{\farc\ spectra of \fsn 2004ie + galaxy ($z = 0.0513$) at $t =
  +5$~days (top), $+19$~days (middle), and $+23$~days (bottom) after maximum.
  The noisy region near $\lambda = 5600$\AA\ is due to the dichroic cutoff of
  the red and blue cameras.} \label{sn83_spec}
\end{figure}

\begin{figure*}[tbh]
\centering
\includegraphics[width=85mm,angle=-90]{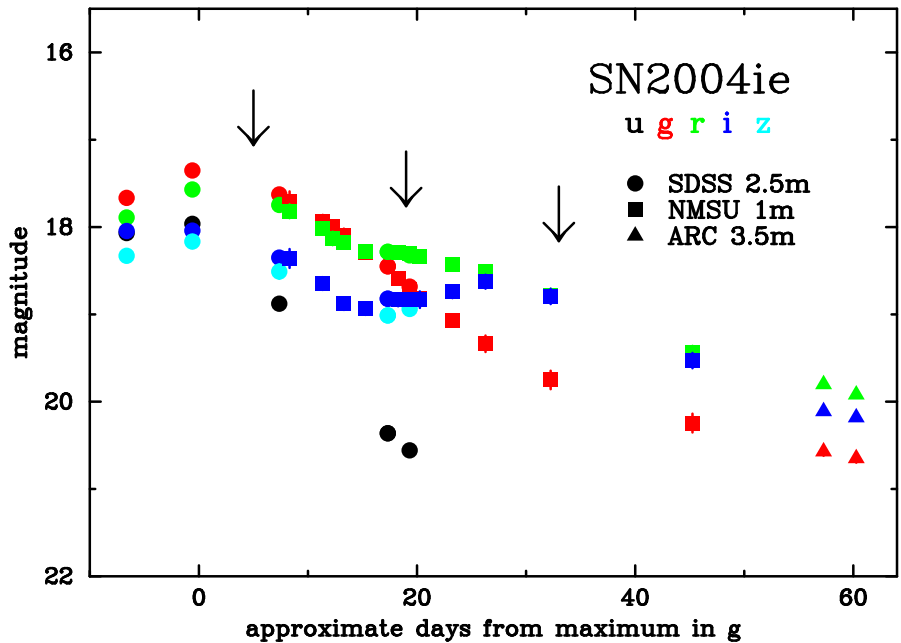}
\caption{Preliminary multi-color light curves of the type Ia \fsn 2004ie ($z =
  0.0513$).  Measurements from the \fsdss\ 2.5m, \fnmsu\ 1m, and the
  \farc\ 3.5m are denoted as filled circles, squares, and triangles,
  respectively.  The three arrows indicate epochs of spectroscopy with the
  \farc\ 3.5m/\fdis. } \label{sn83_lc}
\end{figure*}

\begin{figure}[bth]
\centering
\includegraphics[width=58mm,angle=-90]{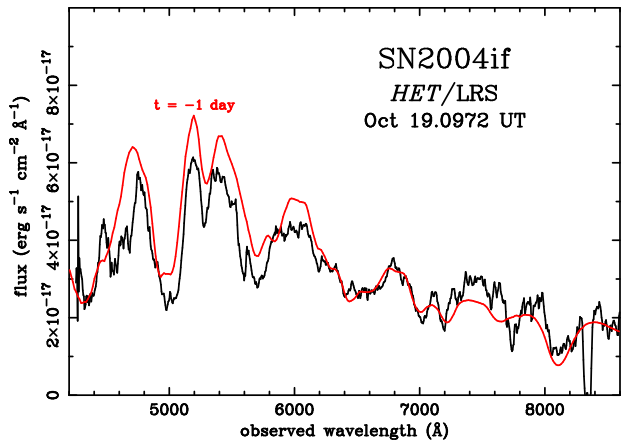}
\caption{Observed spectrum of the type Ia \fsn 2004if ($z = 0.322$) obtained
  with the \fhet\ (black) and the best-matching template spectrum
  (red).} \label{sn25_spec}
\end{figure}

\begin{figure}[htb]
\centering \includegraphics[width=70mm,angle=-90]{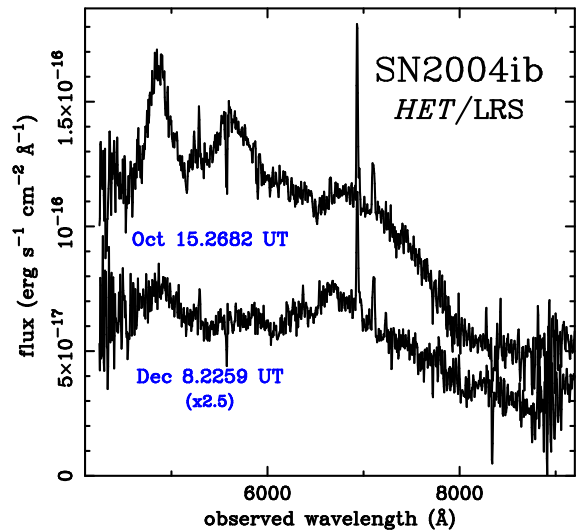}
\caption{Multi-epoch \fhet\ spectra of the type Ib/c hypernova \fsn 2004ib +
  galaxy ($z = 0.0565$).} \label{sn20_spec}
\end{figure}

\begin{figure*}[bth]
\centering
\includegraphics[width=85mm,angle=-90]{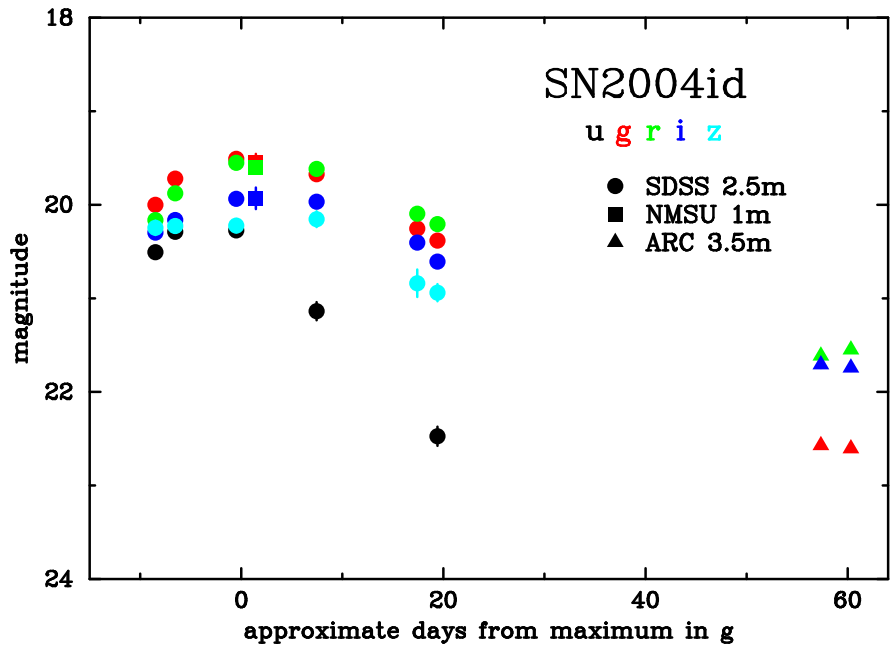}
\caption{Same as in Figure~\ref{sn83_lc} for the type Ia \fsn
  2004id + galaxy ($z = 0.1444$). } \label{sn19_lc}
\end{figure*}

\subsection{Follow-up Spectroscopy}

Spectroscopic observations were planned with Double Imaging Spectrograph
(\dis) on the \arc\ 3.5m telescope at \apo\ and the Low Resolution
Spectrograph (\lrs) on the queue-scheduled 9.2m Hobby-Eberly Telescope (\het)
at McDonald Observatory.  In most cases, relatively bright events with $g \lax
20.6$~mag were targeted with the \arc\ 3.5m and fainter sources were put in
the \het\ queue usually at high priority.  There were a few exceptions,
however, due to scheduling constraints and availability of telescope time on
any given night.  Both observatories are again generally affected by the same
weather pattern.

Spectral classification of type Ia \sne\ is least ambiguous with spectra
obtained near maximum light \cite{filippenko97}, so every attempt was made to
perform the observation within $\sim 2$~weeks of discovery.  This did not
always work in practice, however, particularly for the queue-scheduled \het,
which is affected by conflicts with other competing programs, in addition to
the usual weather and moon constraints.

The redshifts were typically inferred by taking the total measured spectrum
(\sn\ + galaxy) and cross-correlating against a number of template galaxy
spectra using the \iraf\ package \rvsao\ \cite{rvsao}.  When the supernova was
well-separated from the host and no obvious galaxy light was detected, the
spectra were cross-correlated with template \sn\ spectra.  The redshifts
measured from the host galaxy spectra are generally more accurate than
redshifts derived from broad features of the \sn\ spectrum.  In all cases,
however, the two methods agreed to within their statistical errors.

In the Fall 2004 campaign, we were able to spectroscopically confirm 16 type
Ia, 6 type II, and 1 type Ib/c, which are listed in Table~\ref{sn_list} along
with their redshifts.  These include only sources from which we were able to
obtain a \sn\ spectrum.  There are a handful of additional sources for which
the host galaxy redshift is measured, and the \sn\ light curve is consistent
with that of a Ia at this redshift.  These are considered as likely type Ia
\sne, and are not included in Table~\ref{sn_list}.

\section{Light Curves}

Photometric measurements were performed by first registering the template and
search images using point sources near the \sn.  \psf\ magnitudes of the point
sources in the search frame were measured and compared to the catalog
\psf\ magnitudes released in \dr 3 \cite{dr3} to determine the zeropoints in
each of the five colors.  The images were then cross-convolved (template frame
convolved with search \psf; search frame convolved with template \psf) to
minimize systematics \cite{galyam04}.  \psf\ magnitudes were finally measured
on the subtracted images.  Multi-color light curves of two of the best
observed \sn\ Ia are shown in Figures~\ref{sn83_lc} and \ref{sn19_lc}.

Detailed light curve fits are in progress, but a preliminary analysis
indicates that \sn 2004ie is a normal type Ia with $\dm = 1.0$ with very
little host galaxy extinction, consistent with the image shown in the middle
panel of Figure~\ref{snia_images}, which shows the \sn\ near the edge of its
host galaxy.

\section{The Supernova Survey of SDSS II}

The Supernova Survey of \sdss\ II will observe the entire equatorial stripe
82, which covers the coordinate ranges of $-50^\circ < \alpha < +60^\circ$ and
$-1.25^\circ < \delta < +1.25^\circ$, with a cadence of 2 days from the
beginning of September to the end of November for the next three years.  This
is approximately 4 times the size of the Fall 2004 campaign per year.  With
the discovery and confirmation of 16 type Ia \sne, we can estimate that at
least $\sim 16 \times 4 = 64$ \sne\ Ia will be discovered per year with
\sdss\ II.  We expect this to be a lower limit, since (1) during the test run,
our thresholds were not optimized to search for faint events in the redshift
range of $0.2 \lax z \lax 0.35$, which spans a much larger volume, (2)
extending the survey duration from 1.5 months to 3 months results in more than
a factor of two increase in the number of events with well-sampled light
curves, and (3) the spectroscopic follow-up plans for \sdss\ II are also more
extensive.  We expect to request approximately 33 half-nights for
spectroscopic follow-up from the \arc\ 3.5m this Fall.  The \het\ will devote
at least 50 hours of observing time primarily for the $z \gax 0.2$ Ia, and
Ohio State has committed 30 nights per year of \mdm\ 2.4m time for this
project.  In addition, proposals are being submitted to \eso, Subaru, and the
William Herschel Telescope on La Palma.

\begin{figure}[htb]
\centering
\includegraphics[width=70mm,angle=-90]{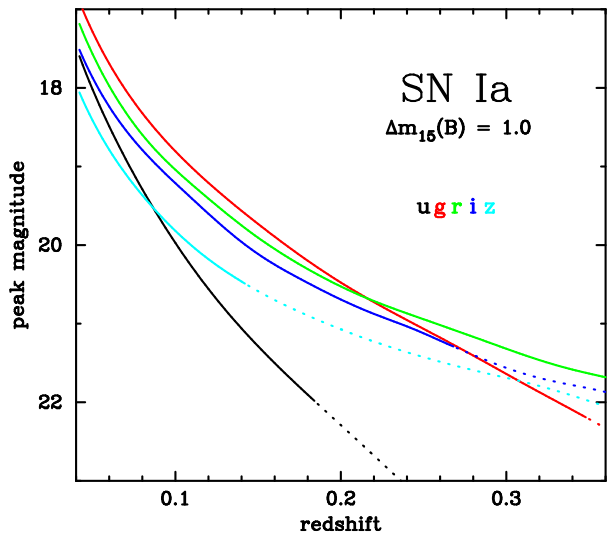}
\caption{Peak magnitudes of a normal type Ia \fsn\ with $\Delta m_{15}(B) =
  1.0$ \cite{phillips93} as a function of redshift calculated using a set of
  template spectra provided by A. Riess (2004, private communication) in each
  of the \fsdss\ filters.  Points on each of the curves that fall below the
  95\% point-source detection repeatability values are shown by dotted
  lines.} \label{peakmag}
\end{figure}

Figure~\ref{peakmag} shows the expected peak \ugriz\ magnitudes for a normal
type Ia with $\Delta m_{15}(B) = 1.0$, where $\Delta m_{15}(B)$ is the
increase in $B$-band magnitude 15 days after peak \cite{phillips93}.  For the
nearby events ($z \lax 0.15$), useful light curves will be acquired in all
five colors, giving better constraints on the reddening.  For more distant
events ($z \gax 0.2$), the $u$ and $z$ magnitudes will likely fall below the
detection limit.

We plan to also run the $i$-band frames through real-time image subtraction in
addition to the $g$ and $r$ frames, and use an improved version of the image
subtraction pipeline for automated candidate selection.  We expect these
changes to result in an increase in the detection efficiency of \sne\ Ia near
the high redshift end of $z \sim 0.3$ without dramatically increasing the
number of candidates for manual inspection.  We also plan to manually perform
photometric measurements on the $u$ and $z$ images of candidates that make
through the cut, and use the full color information for pre-typing.
Photometric redshifts of host galaxies can also be useful for this process.
We continue to optimize our selection criteria using the data acquired in fall
2004.

We also expect to find and measure light curves of $\sim 100$ type II
\sne\ out to a redshift of $z \sim 0.2$ and possibly higher.  The large sample
up to moderate depths will be extremely valuable for studying in detail the
exciting possibility of using type II \sne\ as standardized candles and
applications for cosmology \cite{hamuy02}.  Finally, we note that the large
volume of this survey also provides a unique opportunity to find unusual
\sne\ that are poorly sampled in other surveys.  One such example is the class
of luminous type Ib/c, some of which are thought to be associated with
gamma-ray bursts.  Initial color-typing and photometric redshifts of the host
galaxies can pre-select sources for follow-up spectroscopy in the optical and
possibly in other wavelengths as well.

\bigskip 
\begin{acknowledgments}

The authors would like to thank the \het\ resident astronomers for their quick
cooperative responses, and their continuous effort throughout this run.
{\small MS} thanks Nick Suntzeff for valuable advice.  Funding for the
creation and distribution of the \sdss\ Archive has been provided by the
Alfred P. Sloan Foundation, the Participating Institutions, the National
Aeronautics and Space Administration, the National Science Foundation, the
U.S. Department of Energy, the Japanese Monbukagakusho, and the Max Planck
Society. The \sdss\ Web site is http://www.sdss.org/. The \sdss\ is managed by
the Astrophysical Research Consortium (\arc) for the Participating
Institutions. The Participating Institutions are The University of Chicago,
Fermilab, the Institute for Advanced Study, the Japan Participation Group, The
Johns Hopkins University, the Korean Scientist Group, Los Alamos National
Laboratory, the Max-Planck-Institute for Astronomy ({\small MPIA}), the
Max-Planck-Institute for Astrophysics ({\small MPA}), New Mexico State
University, University of Pittsburgh, University of Portsmouth, Princeton
University, the United States Naval Observatory, and the University of
Washington.

\end{acknowledgments}


\end{document}